\documentclass[%
 reprint,
superscriptaddress,
 amsmath,amssymb,
 aps,
]{revtex4-2}

\usepackage{graphicx}% Include figure files
\usepackage{dcolumn}% Align table columns on decimal point
\usepackage{siunitx} %package for units
\usepackage{bm}% bold math
\usepackage{appendix}
\usepackage{booktabs}
\usepackage{bibunits}
\defaultbibliographystyle{apsrev4-2}
\defaultbibliography{references}
\usepackage[dvipsnames]{xcolor}
\usepackage[
	citecolor=Red,
	linkcolor=Blue,
	filecolor=magenta,
	urlcolor=Blue,
]{hyperref}
\usepackage{physics}
\begin{document}
\begin{bibunit}
\preprint{APS/123-QED}

\title{Dark Optical Trapping of Resonant Transition-Metal Dichalcogenide Particles}

\author{Patrick Illetschek}
\affiliation{University of Vienna, Faculty of Physics, Vienna Center for Quantum Science and Technology (VCQ), Boltzmanngasse 5, A-1090 Vienna, Austria}
\affiliation{Institute for Quantum Optics and Quantum Information (IQOQI) Vienna, Austrian Academy of Sciences, Boltzmanngasse 3, 1090 Vienna, Austria}

\author{Gleb Fedorovich}
\affiliation{Institute of Quantum Electronics ETH Zurich, CH-8093 Zurich, Switzerland}
\affiliation{Department of Physics and Astronomy, Ghent University, Krijgslaan 281, 9000 Gent, Belgium}

\author{Albert Seredin}
 \affiliation{Emerging Technologies Research Center, XPANCEO, Dubai Investment Park 1, Dubai, 00000 UAE}

\author{Gleb Tselikov}
 \affiliation{Emerging Technologies Research Center, XPANCEO, Dubai Investment Park 1, Dubai, 00000 UAE}

 \author{Valentin S. Volkov}
 \affiliation{Emerging Technologies Research Center, XPANCEO, Dubai Investment Park 1, Dubai, 00000 UAE}
 
 \author{Nikolai Kiesel}
\affiliation{University of Vienna, Faculty of Physics, Vienna Center for Quantum Science and Technology (VCQ), Boltzmanngasse 5, A-1090 Vienna, Austria}

\author{Markus Aspelmeyer}
\affiliation{University of Vienna, Faculty of Physics, Vienna Center for Quantum Science and Technology (VCQ), Boltzmanngasse 5, A-1090 Vienna, Austria}
\affiliation{Institute for Quantum Optics and Quantum Information (IQOQI) Vienna, Austrian Academy of Sciences, Boltzmanngasse 3, 1090 Vienna, Austria}

\author{Mihail Petrov}
 \affiliation{Emerging Technologies Research Center, XPANCEO, Dubai Investment Park 1, Dubai, 00000 UAE}

\author{Anton V. Zasedatelev}
\email{anton.zasedatelev@aalto.fi}
\affiliation{University of Vienna, Faculty of Physics, Vienna Center for Quantum Science and Technology (VCQ), Boltzmanngasse 5, A-1090 Vienna, Austria}


\date{\today}% It is always \today, today,

\begin{abstract}
Mitigating recoil events and minimizing optically induced heating are central challenges in the precise control and cooling of macroscopic particles. To overcome this, we propose trapping resonant dielectric particles for applications in ultra-high vacuum (UHV) levitodynamics. Contrary to other approaches, where suppressing the parasitic resonant scattering was achieved in a standing wave geometry, here we propose a single beam geometry in a dark trap regime. As a promising material platform, we focus on a class of transition-metal dichalcogenide (TMD) particles with high polarizability, characterized by refractive indices in the range $3.7$--$4.8$ and densities up to $9.3~\mathrm{g\,cm^{-3}}$. Using full Mie theory, we identify a range of TMD particle radii that support stable axial and radial magnetic quadrupole trapping in a bottle-beam configuration. We predict that for WS$_2$ particles with a mass of $\num{0.5e12} \si{ amu}$, one can expect suppression of the scattering rate relative to the mechanical frequency down to $\Gamma/\Omega \simeq 0.02$. This corresponds to a coherence time extended by approximately three orders of magnitude compared with silica particles of the same mass trapped in conventional bright optical traps at UHV.  Combined with significantly reduced internal heating, remaining well below the melting point of the material, dark trapping of resonant TMD macroscopic particles emerges as a promising platform for exploring quantum physics with large masses.

\end{abstract}

\maketitle

Optical levitation of nano- and micro-particles in vacuum has developed into a versatile platform for exploring both fundamental physics and sensing applications \cite{levitodynamics}. In particular, levitated particles in ultra-high vacuum (UHV) are extremely well isolated from their environment, forming low-dissipation mechanical oscillators, %with a Q-factor exceeding $10^{8}$ \cite{millen2020optomechanics,dania2024ultrahigh}
whose motion can be continuously monitored and steered at the quantum level \cite{lorenzocontrol,tebbenjohanns2021quantum,kamba2023nanoscale}. %In recent years, rapid progress in optomechanics with optically trapped masses has pushed these systems toward the ground state cooling inside optical cavities \cite{groundstate} and in a free space \cite{lorenzocontrol,tebbenjohanns2021quantum}. 
Superior control over the motion of optically trapped particles, together with active feedback schemes, enables the generation of ponderomotive squeezing of light \cite{magrini2022squeezed,squeezing_eth} and, more recently, squeezing of the mechanical motion of a particle below the standard quantum limit \cite{squeezing_aikawa}. Beyond single-particle control, emerging approaches also enable the engineering of interactions between multiple levitated particles \cite{rieser2022tunable,reisenbauer2024non,vijayan2023scalable} and exploration of collective quantum phenomena, e.g. entanglement generation between large masses optically trapped in UHV. \cite{rudolph2022force,winkler2025steady,poddubny2024nonequilibrium}.

At the same time, the optical fields used for trapping and measuring inevitably introduce measurement backaction through photon scattering. Each scattered photon transfers momentum to the particle, leading to photon-recoil heating of the centre-of-mass motion \cite{recoilrate}. Moreover, photon recoil in optical traps in UHV conditions becomes a dominant source of decoherence and limits the time over which coherent quantum dynamics can be maintained. Several strategies have therefore been explored to mitigate recoil heating and reduce optical backaction. These include trapping particles at an intensity minimum (dark trapping) with active stabilization \cite{electrodes}, cavity-based approaches that modify the optical density of states and suppress scattering \cite{highpurity_optomechanics,cavity_squeezing}, and the use of reflective boundaries or engineered optical environments to control light scattering \cite{backaction_supress,reflecitve_supression}.

%===========Figure=====================
\begin{figure}[t]
\includegraphics[width=0.8\linewidth]{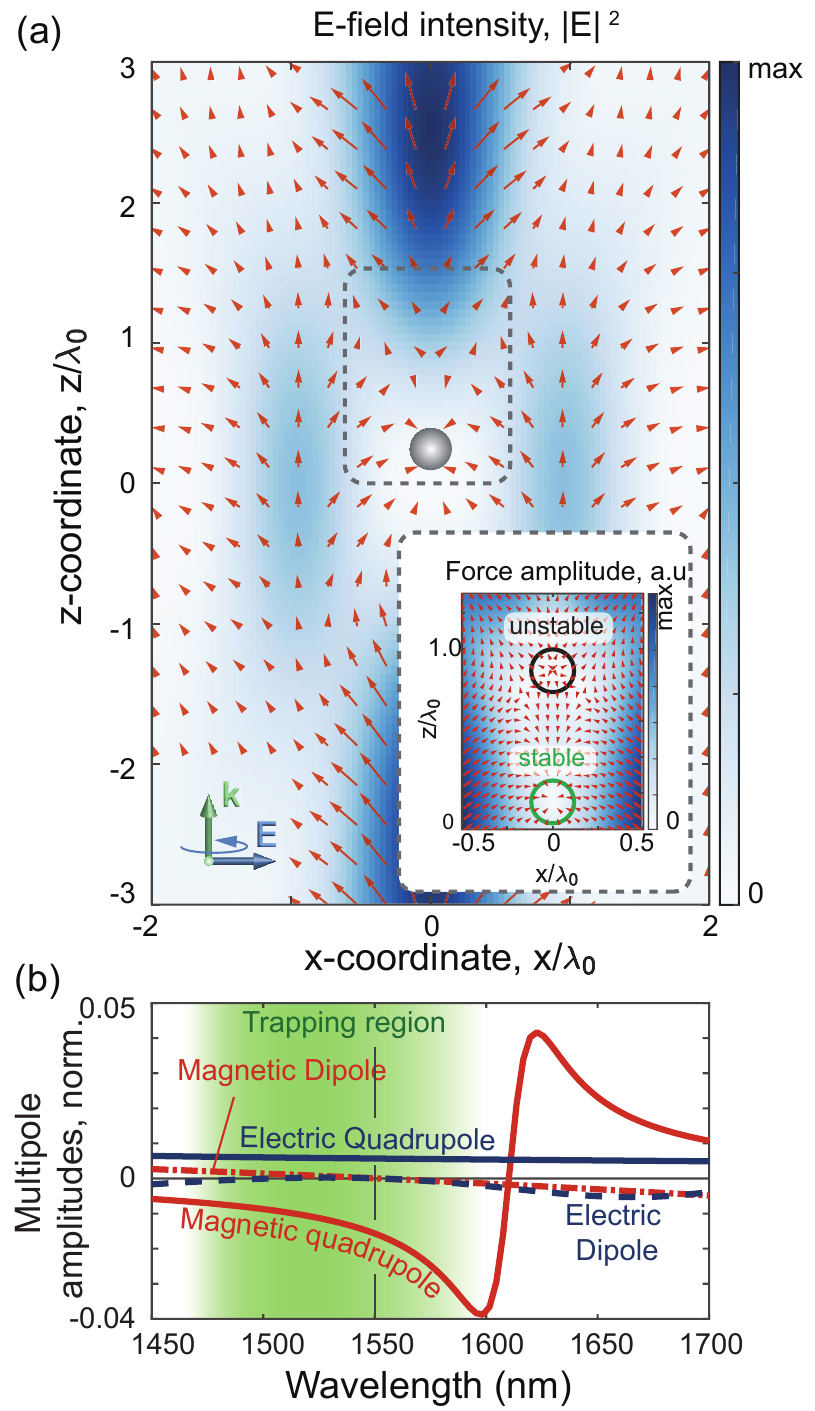}
\caption{\label{fig1} \textbf{The dark trap.}
(a) Normalized EM field intensity in the ($xz$) plane of the bottle beam formed by two circularly polarized Laguerre Gauss (LG) beams with a $\pi$ phase shift at $\lambda_0 = 1550$\si{ nm}; orange arrows indicate force vectors. The beam propagates in the +z direction. The inset shows the force amplitude near the origin, revealing the presence of both a stable and an unstable trapping position. 
(b) Multipole amplitudes of the field scattered by the particle near the magnetic quadrupole (MQ) resonance. The green-shaded region marks the wavelength range of stable trapping, while the dashed black line shows the trapping beam wavelength.}
\vspace{-7pt}
\end{figure}
%===========Figure=====================

Alternatively, one can control the scattering properties of particles to achieve a particular functionality. In this perspective, high-index resonant nanoparticles are very promising candidates for applications in optomechanics due to their light scattering properties governed by Mie resonances. While their stable control is still an open problem in optomechanics, they were recently proposed as a promising alternative platform for levitodynamics studies \cite{metaatoms}. However, the common single-beam configuration can not be directly implemented due to resonantly enhanced parasitic pressure force acting on Mie particles, and more sophisticated trapping approaches involving standing wave trapping should be used \cite{juan_cooperatively_2017, toftul_optical_2025, mao_switchable_2025}. In this work, we propose a dark trap geometry with a single-sided excitation in a bottle beam configuration that can be achieved by interfering two co-propagating beams \cite{hologrambeam} or with a single input beam by using methods of spatial light modulation \cite{alpmann_holographic_2012, bottlebeamgeneration,Si_saphire_metasurface_bb}.      

Dark trapping in optical intensity minima has been successfully implemented for neutral atoms using blue-detuned bottle-beam traps \cite{Rubidium_trap}. Similar trapping configurations have also been demonstrated for highly excited Rydberg atoms, where the spatial structure of the optical field allows confinement near the intensity minimum \cite{Ry_bottlebeam}. Concepts of trapping in intensity minima have been further extended to larger objects, for example, a dielectric microparticle trapped in the dark region of a bottle beam in liquid \cite{thiagoexperiment,thiagotheory}. In that case, effective negative polarizability arises from the strong refractive index contrast between the particle and the surrounding medium, allowing stable trapping at the intensity minimum. Alternatively, dark beam schemes can be utilized for trapping particles in gaseous media, assisted by the photophoretic effect \cite{shvedov_robust_2011, alpmann_holographic_2012}.  Implementing similar dark trapping schemes for massive particles in UHV is significantly more challenging.

Pushing trapped particles in UHV toward Planck-scale masses ($m_P\simeq 22\mu g$) opens a route to table-top tests of quantum gravity, where accessing regimes of gravity-mediated entanglement, gravitational decoherence, and possible breakdown of quantum superposition requires maximizing mass while preserving coherence \cite{carney2019tabletop,bose2025massive}. However, increasing mass simply by enlarging optically trapped dielectric particles is fundamentally limited by photon scattering and recoil heating, which affects stability and coherence time, motivating high-density materials and new configurations that enable trapping larger masses with reduced photon recoil. In this perspective, resonant high-index, high-density particles based on transition-metal dichalcogenide (TMD) materials provide a promising platform \cite{tmds_review} for realizing dark trapping of massive particles in a bottle-beam configuration since the fabrication of TMD nanoparticles has been recently developed \cite{gleb_tmds, gleb2025tunable}. These materials possess a large refractive index \cite{tmd_high_index} and a rich spectrum of internal degrees of freedom, which have already enabled a variety of photonic applications such as harmonic generation \cite{harmonic_generation} and quasi-phase-matched up- and down-conversion \cite{rozema} among others \cite{wu2024synthesis}. The high refractive index of TMDs gives rise to strong Mie resonances, as we show here for sub-wavelength particles, allowing control over the magnitude and even the sign of optical forces and torques acting on a particle \cite{afridi2026controlling, toftul_optical_2025, toftul_nonlinearity-induced_2023}. This principle has recently been demonstrated experimentally in a metasurface platform, where engineered multipolar modes enabled control over the sign of the optical force and attraction toward the intensity minimum of a standing wave \cite{afridi2026controlling}, in close analogy to blue-detuned atom trapping. Alternatively, isolating a single magnetic resonance with structured beams enables stable trapping of silicon particles \cite{li_optical_2025}. 

In this work, we theoretically demonstrate that coherent quantum dynamics is achievable with TMD particles in a dark optical trapping configuration at UHV conditions. Specifically, we investigate the optical forces, decoherence, and heating in a bottle beam configuration, which represents a structured optical field with near-zero intensity at its centre, surrounded by high intensity regions. These can be generated by the interference of two overlapping Gaussian beams \cite{atomtrapinbottlebeam,bottlebeamgeneration}, through metasurfaces with the appropriate phase profiles \cite{bottlebeamgeneration,Si_saphire_metasurface_bb}, by spatial light modulation \cite{hologrambeam}, or other methods.
Unlike dark trapping in a standing-wave configuration, bottle-beam traps can be realized using a single input beam \cite{bottlebeamgeneration}, and therefore do not rely on maintaining a stable phase relation between counterpropagating fields over the trapping region, which can otherwise limit motional coherence through phase diffusion. The bottle-beam geometry is thus expected to be less sensitive to this mechanism, making momentum diffusion due to photon recoil the dominant fundamental source of decoherence.
For simplicity, we follow the approach outlined in Ref. \cite{thiagotheory}, where the interference of two Laguerre Gauss (LG) beams was suggested: a loosely focused ($n=0,l=0$) beam with a waist of $w_1= \num{1.59e-6}$ m and a tightly focused  ($n=1,l=0$)  beam with a waist of $w_2 = \num{8.66e-7}$ m, having a relative phase relation of $\pi$ at $\lambda_0 = 1550$ nm. Both beams are circularly polarized. 
The axial cross-section of the EM-field intensity distribution of the bottle beam is shown in Fig.~\ref{fig1} (a) by colour. One can see the formation of a zero-intensity region in the centre of the beam. Its width can be controlled by tuning the relative powers and waists of the two LG beams. 
%===============TABLE======================================
\begin{table}[h!]
\centering
\caption{Index of refraction $n$, extinction coefficient $k$, and density of different particle materials for optical trapping at $\lambda = 1550$ \si{nm}.}

\begin{tabular}{l c @{\hspace{1cm}} c c} % added 1cm space between n and k
\toprule
Material & $n$  & $k$  & Density (g/cm$^3$) \\  
\midrule
WS$_2$ & 3.79 & $<$ \num{2.83d-10} \cite{WS2n}& 7.50 \cite{densityws2} \\
MoS$_2$ & 3.76  & $<$ \num{2.37d-11} \cite{indexmos2}& 5.06 \cite{handbook_of_chem} \\
Si & 3.48  & \num{5.30d-11} \cite{Siindex} & 2.33 \cite{metaatoms}\\
SiO$_2$ & 1.44 & \num{2.55d-16} \cite{sio2index} & 2.20 \cite{metaatoms}\\
\bottomrule
\label{densitytable}
\end{tabular}
\end{table}
%===============TABLE======================================

Hereafter, we focus mainly on the results for a tungsten disulphide (WS$_2$) particle due to its comparatively higher density, while other TMD materials are studied in detail in the Supplementary materials. WS$_2$ possesses a high refractive index of $n(\lambda_0) = \num{3.79} + \num{2.83e-10}i$ and a density of $\rho = {7.5}$ g/cm$^3$, placing it among the higher-density semiconductor materials  (exceeding the density of silica particles by more than 3 times, see Table~\ref{densitytable}). As a TMD, bulk WS$_2$ is an indirect band-gap semiconductor exhibiting low absorption losses in the infrared, together with a large refractive index and high material density \cite{wu2024synthesis}.

Here, we consider a spherical particle of radius 300 nm at the light wavelength $\lambda_0=1550$ nm. Its optical response is governed by low-order Mie resonances \cite{GustavMie1908}. Consequently, the electromagnetic field of the incident and scattered waves can be decomposed into a multipolar series \cite{scattmisch}:

\begin{align}
\mathbf{E}_{\text{in (sca)}}(\mathbf{r}) 
   = \sum_{n,m}^{\infty}\sum_{m=-n}^{n}
      \Big[& a_{nm}(p_{nm})\,\mathbf{M}_{nm}(k\mathbf{r})
           +\nonumber \\
           &+b_{nm}(q_{nm})\,   \mathbf{N}_{nm}(k\mathbf{r}) \Big], 
\label{Efield}
\end{align}

where $a_{nm}(p_{nm})$ and $b_{nm}(q_{nm})$ are the incoming (scattered) magnetic and electric multipole amplitudes with corresponding total angular momentum (or multipole order) $n$ and its projection on the z-axis (or azimuthal order) $m$. $\vb{M}_{nm}$ and $\vb{N}_{nm}$ are the magnetic and electric vector spherical harmonics (VSH) with proper radial dependence for the incident and scattered waves \cite{scattmisch, toftul_nonlinearity-induced_2023}, and $k$ is the wavenumber of the incident wave. 

The imaginary parts of the scattered field amplitudes for the particle placed in the centre of the bottle beam are shown in Fig.~\ref{fig1} (b).  One can see that since the particle is placed in the field node, both electric (ED) and magnetic (MD) components are suppressed. While the electric quadrupolar (EQ) response is also relatively small, the main contribution is attributed to the resonant magnetic quadrupole (MQ)  response. We should stress that in this spectral region, ED, MD and MQ components are on the blue side of the resonance, thus they add to the gradient force by pushing the particle out of the high-intensity region. The EQ contribution, on the contrary, tends to pull the particle away from the field node. This competition, along with the enhanced scattering force close to MQ resonance, forms the essence of the force balance. Our analysis shows that although the magnetic quadrupole strongly influences the field near the beam waist and contributes to the formation of the dark trapping region, the equilibrium position and trap stiffness are determined by the interference of all four multipole contributions. The analysis of purely MQ contribution is provided in the Supplementary materials, showing good qualitative but not quantitative agreement. 
%===========Figure2=====================
\begin{figure}[t]
\includegraphics[width=1\linewidth]{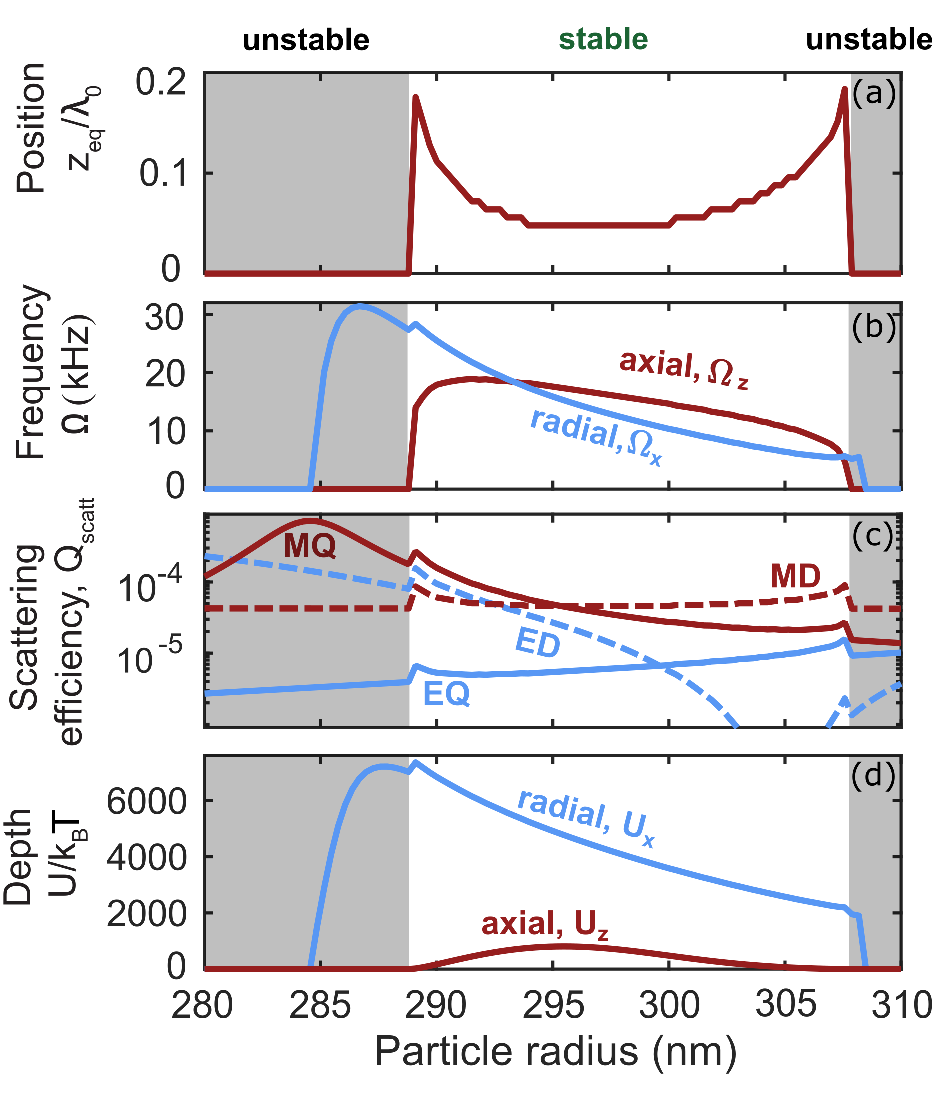}
\caption{\label{fig2} \textbf{Stable trapping conditions}. (a) Equilibrium position $z_{eq}$ versus particle radius, normalized to $\lambda_0 = 1550$ nm.
(b) Trap frequencies: axial $z$ (red) and in-plane, radial $x,y$ (blue) for $P={0.80}$ W.
(c) Scattering efficiencies $Q_{\text{scatt}}$ of multipoles at $z_{eq}$ for different particle radii: Magnetic quadrupole (MQ), Magnetic Dipole (MD), Electric Dipole (ED), Electric Quadrupole (EQ).
(d) Trap depth in units of $k_B T$ ($T=300$ K), along radial $x,y$ (blue) and axial $z$ (red) directions, for different particle radii.
The shaded regions indicate particle radii where no 3D stable trapping is found at $\lambda_0 = 1550$ nm.}
\vspace{-7pt}
\end{figure}
%==============Figure 2===================

 The distribution of the total optical force across the beam is visualized with arrows in Fig.~\ref{fig1} (a). Here, the force accounts for all multipolar components and is calculated with the Optical Tweezer Toolbox \cite{ott} by integrating the Maxwell stress tensor over a surface surrounding the particle \cite{farsund1996force}. The inset shows the existence of an unstable and a stable trapping region. One can see that the particle of $R=300$ nm is pushed away from the regions of high intensity to the stable point $z=z_{\text{eq}}$. Its dependence on the particle radius is shown in Fig.~\ref{fig2}(a). The shift of $z_{\text{eq}}$ to the positive z-direction results from the optical forces governed by the Mie resonances and the scattering force imposing radiation pressure on the particle. The range of particle radii for which stable trapping is achieved is delineated by the white regions, while the gray shaded regions indicate the absence of stable trapping. The stable region roughly corresponds to the region of negative polarizability $\text{Re}(\alpha_{\text{MQ}})<0$, where the optical MQ gradient force dominates over the other force components. Detuning to the red side is limited by the MQ resonance, as the scattering efficiency at the top of the MQ resonance is resonantly increased. Detuning to the blue side is constrained by the increasing EQ and ED contributions, whose positive polarizabilities push the particle toward the field maxima. In general, stable dark trapping can be achieved for a much broader range of particles provided that the wavelength of the trapping laser can be tuned over a sufficiently wide range.

For the found stable position on the beam axis, we plot the axial (red curve) and radial (blue curve) frequency of the trap as a function of particle radius in Fig.~\ref{fig2} (b) for laser beams with total power $P_0 = {0.8}$ W. We provide the trap frequency only for the region where stable trapping (white area) exists. It is also important to emphasize that, in contrast to the case of bright tweezers, the dark trap has a higher axial frequency than radial. This behaviour is reminiscent of standing wave optical traps \cite{hollow_core_trap}. In Fig.~\ref{fig2} (c), we plot the scattering efficiency $Q_{\text{sca}}(z_{\text{eq}})=\sum_{nm}(|p_{nm}|^2+|q_{nm}|^2)$  of the particle decomposed over the multipolar components. One can see that while the MQ component dominates for smaller particle size, the other components are non-zero since the particle is shifted from the beam centre. We also should note that for the unstable region, we provide the scattering intensity calculated at the beam centre $z=0$ since no stable point along $z$-axis exists. That explains the jump of $Q_{\text{sca}}$ at the border of stable/unstable regions. Finally, in Fig.~\ref{fig2} (d), the axial and radial trap depth $U$ in units of $k_BT$ ($T= 300$ {K}) is plotted. The radial trap depths are obtained from the effective potential derived by integrating the corresponding force components along the respective axes. The axial trap depth is determined from a local harmonic approximation based on the axial stiffness at $\text{z}_{\text{eq}}$. This shows that the dark trap provides feasible experimental conditions for three-dimensional stable trapping of a TMD sub-micron particle.

One of the key parameters defining the trapping efficiency is the photon scattering rate of the particle, $\Gamma =  \omega P_{\mathrm{sca}}/(5m c^2\Omega_0)$, where $\omega$ is the frequency of the incident beam, $\Omega_0$ the frequency of the trap and $P_{sca} = P_{inc} \cdot Q_{sca}$ the scattered power \cite{recoilrate}. Photon recoil introduces momentum kicks that act as a random force driving the particle \cite{recoilrate,delocalization}. 
The ratio $\Gamma/\Omega$ therefore quantifies the relative strength of recoil-induced diffusion and determines the number of coherent oscillations for the particle in a trap. Other mechanisms, such as phase-diffusion, can also limit the number of coherent oscillations; however, these are expected to be less severe in the present single-sided bottle-beam geometry than in standing-wave configurations.

\begin{figure}[t]
\includegraphics[width=0.95\linewidth]{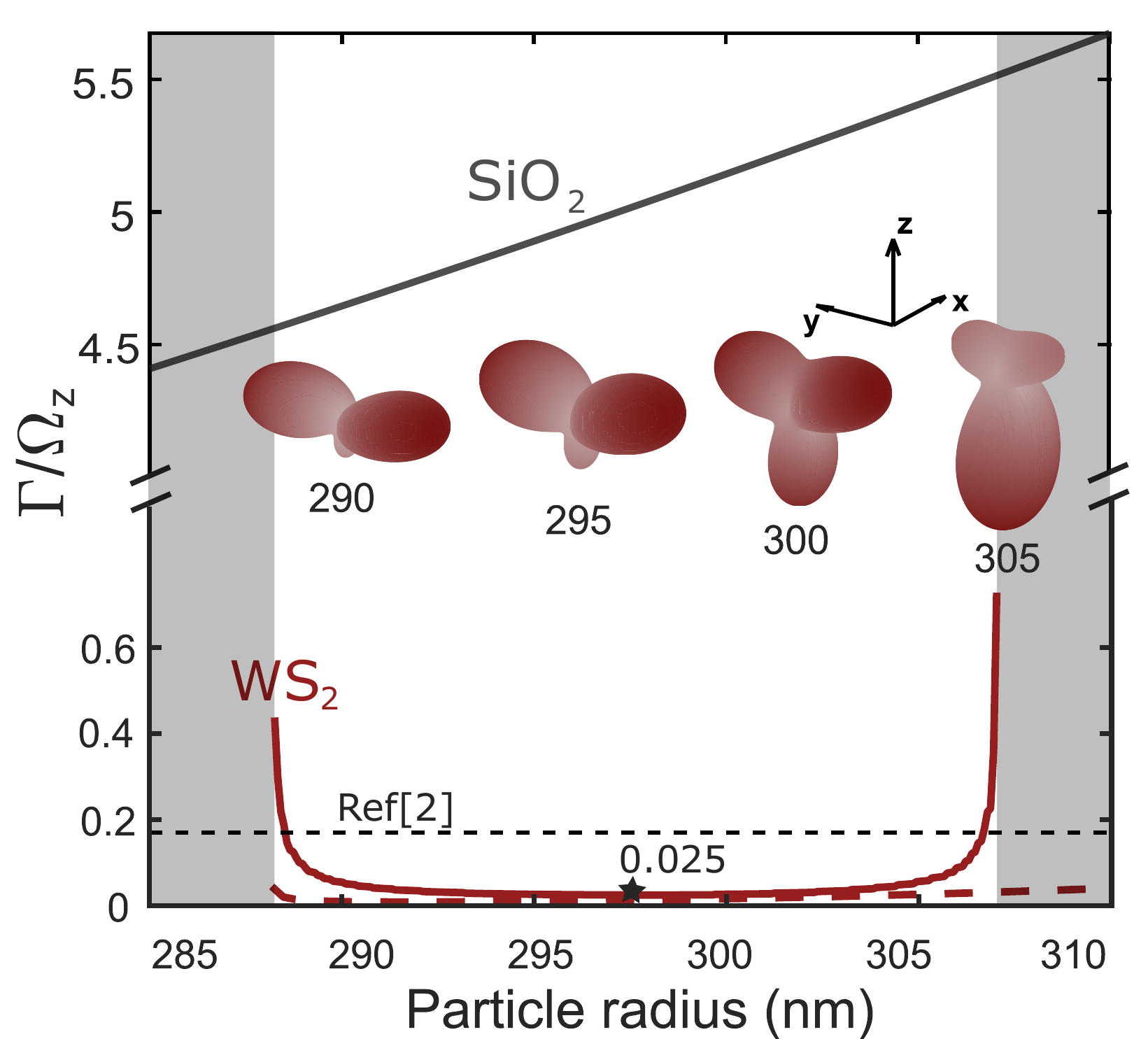}
\caption{\label{fig3} \textbf{Photon recoil in the dark trap}. 
Ratio of photon recoil rate to trap frequency, $\Gamma/\Omega_z$, versus particle radius for a WS$_2$ particle in the dark trap (red solid line), with a compensating force (red dashed line), for a SiO$_2$ particle of the same radius (black solid line) and of radius 71.5 nm (black dashed line) \cite{lorenzocontrol}) trapped in a conventional bright trap. Overlaid are the scattered intensity patterns for particles of different radii.}
\vspace{-7pt}
\end{figure}

In Fig.~\ref{fig3}, we benchmark the photon recoil of WS$_2$ particle in the dark trap (solid red line) against a conventional SiO$_2$ particle of the same radius in a bright trap (black solid line). The black dotted line indicates the ratio $\Gamma/\Omega_z = 0.18$, which corresponds to a SiO$_2$ particle with $R = {71.5}$ nm ($m = {1.7\cdot 10^9}$ amu), as these are typical particles used in optical trapping experiments  \cite{lorenzocontrol}. For a trapped WS$_2$ particle of radius 298 nm ($m =0.5 \cdot 10^{12}$, amu), we obtain $\Gamma/\Omega_z = 0.025$. This corresponds to nearly an order-of-magnitude reduction in recoil-induced decoherence compared to a conventional SiO$_2$ particle in a bright trap, despite the latter being approximately 300 times lighter. For particles of the same radius, the reduction exceeds two orders of magnitude. 

%===========Figure=====================
%===========Figure=====================

To compensate for the radiation pressure shift in the equilibrium position, one can introduce a restoring force, for example, by applying an electrostatic force to the charged particle using a pair of electrodes \cite{electrodes}. This further allows for the expansion of the stability region and keeps photon recoil at the minimal level $\Gamma/\Omega_z = 0.0092$, as shown in Fig.~\ref{fig3} (red dashed line). Overlaid in Fig.~\ref{fig3} are the far-field scattered intensity diagrams for different particle radii (noted below each diagram), which are directly calculated from the scattered field provided by Eq.~\ref{Efield}.\\

\begin{figure}[t]
\includegraphics[width=0.95\linewidth]{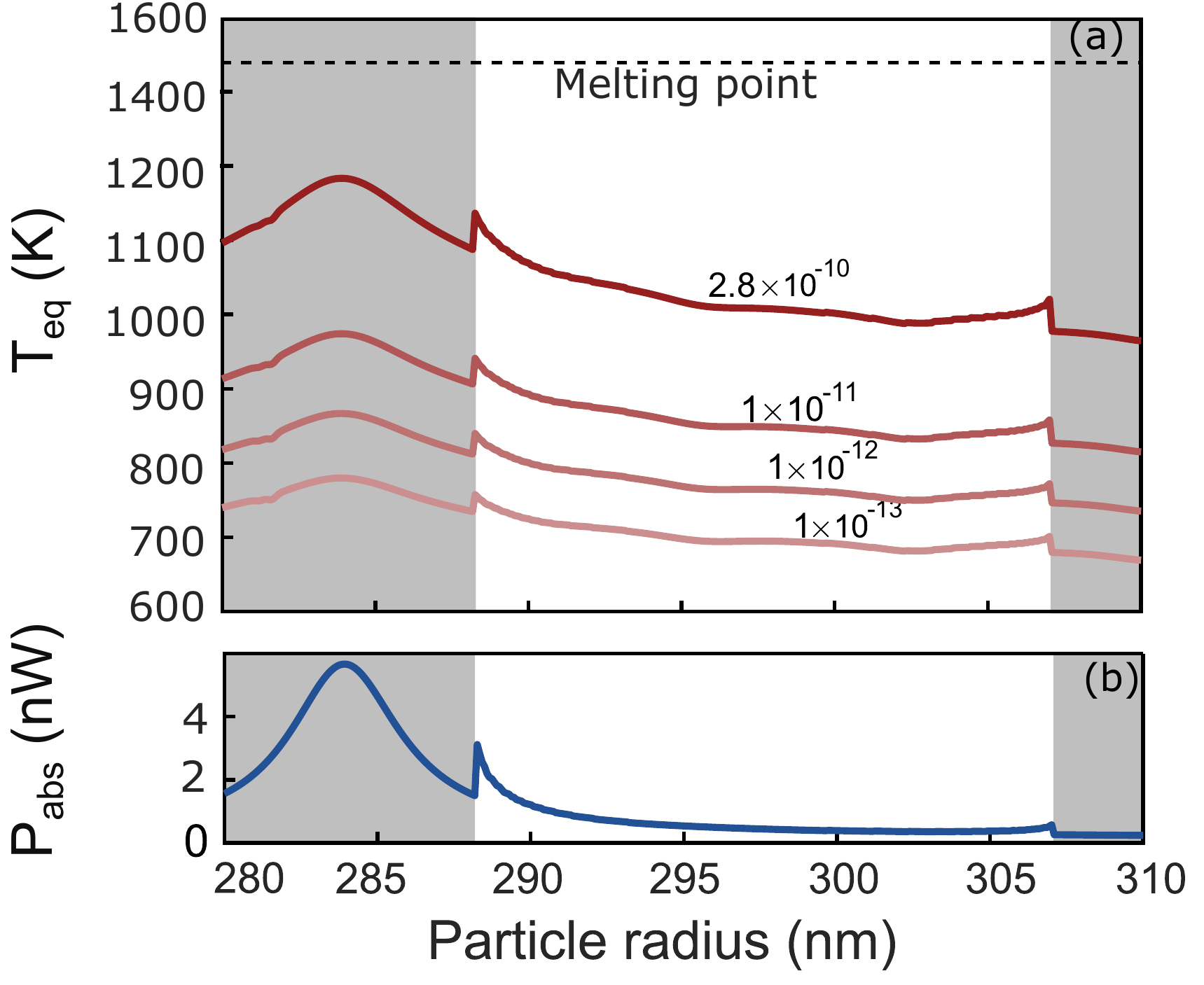}
\caption{ \textbf{Internal heating and thermal emission}. (a) Equilibrium temperature $T_{\text{eq}}$ versus particle radius for different extinction coefficients $ k$. The black dotted line indicates the melting point of WS$_2$ at 1520 K.
(b) Absorbed power of the trapped nanoparticle. Shaded regions indicate particle radii where stable trapping is not achieved.}
\label{fig4}
\vspace{-7pt}
\end{figure}

In addition to photon recoil, laser-induced internal heating presents a major challenge for quantum experiments with increasingly large particles in UHV. Even weak absorption in bright traps can cause a substantial temperature rise, as blackbody radiation is the primary \textit{(and often the only)} thermalization mechanism. For example, silicon particles at low pressures can escape from the trap due to heating dominated by two-photon absorption \cite{silion_absorption}. Increased internal temperature also leads to decoherence of center-of-mass motion via blackbody emission \cite{blackbody_decoherence}, even if the photon recoil is completely suppressed. Here, we calculate absorption and blackbody radiation for resonant particles in the dark trap in UHV and determine its equilibrium temperature $T_{\text{eq}}$, which constrains the optical power of the trap, particle size, and material. Figure~\ref{fig4}(a) shows the equilibrium temperature $T_{\text{eq}}$ of the particle in the dark trap, obtained under the assumption that cooling occurs solely via radiative processes \cite{metaatoms, kallel_temperature_2017}. Importantly, resonant TMD particles exhibit a more pronounced spectral dependence in their emission, in contrast to SiO$_2$ particles. In fact, thermal radiation is strongly enhanced at specific wavelengths corresponding to the particles' Mie resonances. This effect is accounted for by weighting the spectral radiance with the absorption cross-section and integrating over the relevant spectral range to get the radiated power of the trapped particle. Heating is dominated by absorption of the trapping laser proportional to the absorption efficiency $Q_{\text{abs}}$ and thus the absorbed power (dark blue line in Fig.~\ref{fig4}(b)).

%===========Figure=====================
%===========Figure=====================

Despite operating in a dark trap, residual optical fields still induce weak absorption and heating. Figure ~\ref{fig4}(a) presents the dependence of the equilibrium temperature $T_{\text{eq}}$ for different values of the imaginary part of the refractive index, $k$. In the spectral range above {938} nm, optical losses in TMD materials become extremely low, to the extent that standard experimental techniques struggle to reliably resolve $k$ \cite{indexmos2}, resulting in significant uncertainty in its exact value. To account for this, we consider a range of values spanning $k = 10^{-13}$ to $k = 10^{-10}$. Nevertheless, the equilibrium temperature in the dark trap remains well below the melting point of WS$_2$ (black dotted line in Fig.~\ref{fig4}(a)), which is approximately {1520} K \cite{densityws2}. The laser pump power was set to $P_0=0.8 \si{ W}$. Importantly, owing to the large band gap of TMD materials, two-photon absorption at telecom wavelengths is strongly suppressed, in contrast to silicon particles at {1550} nm, where nonlinear absorption can contribute significantly to internal heating. One should also note that here we did not consider thermo-optical effects induced by optical heating \cite{Zograf2021} such as resonance position shift, which can result in an abrupt change of the optical properties of the nanoparticle \cite{Duh2020, ryabov_nonlinear_2022, nishida_optical_2023, nishida_all-optical_2024}.

We investigate dark optical trapping of resonant transition-metal dichalcogenide particles in ultra-high vacuum. Using Mie theory, we identify a size range where stable trapping is achieved in the magnetic quadrupole (MQ) regime, providing full 3D confinement.
TMD particles combine high density \textit{(up to 5 times the density of SiO$_2$) }with resonant optical response, enabling access to larger masses while suppressing photon recoil. For a WS$_2$ particle of mass $\num{0.5e12},\si{amu}$, we obtain $\Gamma/\Omega_z \simeq 0.025$, corresponding to 3 orders-of-magnitude reduction compared to SiO$_2$ particles of the same mass in bright traps ($\Gamma/\Omega_z = 20.4$). Together with reduced internal heating well below the melting point, resonant TMD particles provide a promising platform for levitating larger masses with low motional decoherence. 
It has been pointed out that large recoil implies potentially large measurement rates, which can improve ground state cooling speed and may enable strong coupling in free space \cite{metaatoms,maurerdetection}. On the other hand, exploiting sculpted optical potentials for non-linear quantum state engineering requires low localization rates, i.e. low recoil heating as evident in the particle size limit for exploiting the optical cubic nonlinearity for matter-wave experiments with large masses \cite{neumeier2207fast}. Future research will show if exploiting Mie resonances is a viable tool to extend such a protocol in a regime of still larger masses.

\section*{Acknowledgments}

We acknowledge support from the European Union’s Horizon 2020 research and innovation program under Grant Agreement No. 951234, and from the Marie Skłodowska-Curie grant LOREN, Grant Agreement No. 101030987, for A.V.Z.

\clearpage
\putbib
\end{bibunit}

\clearpage
\onecolumngrid

\begin{bibunit}

\renewcommand{\thefigure}{S\arabic{figure}}
\renewcommand{\thetable}{S\arabic{table}}
\renewcommand{\theequation}{S\arabic{equation}}

\renewcommand{\theHfigure}{supp.figure.\arabic{figure}}
\renewcommand{\theHtable}{supp.table.\arabic{table}}
\renewcommand{\theHequation}{supp.eq.\arabic{equation}}

\setcounter{figure}{0}
\setcounter{table}{0}
\setcounter{equation}{0}

\makeatletter
\setcounter{affil}{0}
\makeatother

\title{Supplementary Material for\\
Dark Optical Trapping of Resonant Transition-Metal Dichalcogenide Particles}

\author{Patrick Illetschek}
\affiliation{University of Vienna, Faculty of Physics, Vienna Center for Quantum Science and Technology (VCQ), Boltzmanngasse 5, A-1090 Vienna, Austria}
\affiliation{Institute for Quantum Optics and Quantum Information (IQOQI) Vienna, Austrian Academy of Sciences, Boltzmanngasse 3, 1090 Vienna, Austria}

\author{Gleb Fedorovich}
\affiliation{Institute of Quantum Electronics ETH Zurich, CH-8093 Zurich, Switzerland}
\affiliation{Department of Physics and Astronomy, Ghent University, Krijgslaan 281, 9000 Gent, Belgium}

\author{Albert Seredin}
 \affiliation{Emerging Technologies Research Center, XPANCEO, Dubai Investment Park 1, Dubai, 00000 UAE}

\author{Gleb Tselikov}
 \affiliation{Emerging Technologies Research Center, XPANCEO, Dubai Investment Park 1, Dubai, 00000 UAE}

 \author{Valentin S. Volkov}
 \affiliation{Emerging Technologies Research Center, XPANCEO, Dubai Investment Park 1, Dubai, 00000 UAE}
 
 \author{Nikolai Kiesel}
\affiliation{University of Vienna, Faculty of Physics, Vienna Center for Quantum Science and Technology (VCQ), Boltzmanngasse 5, A-1090 Vienna, Austria}

\author{Markus Aspelmeyer}
\affiliation{University of Vienna, Faculty of Physics, Vienna Center for Quantum Science and Technology (VCQ), Boltzmanngasse 5, A-1090 Vienna, Austria}
\affiliation{Institute for Quantum Optics and Quantum Information (IQOQI) Vienna, Austrian Academy of Sciences, Boltzmanngasse 3, 1090 Vienna, Austria}

\author{Mihail Petrov}
 \affiliation{Emerging Technologies Research Center, XPANCEO, Dubai Investment Park 1, Dubai, 00000 UAE}

\author{Anton V. Zasedatelev}
\email{anton.zasedatelev@aalto.fi}
\affiliation{University of Vienna, Faculty of Physics, Vienna Center for Quantum Science and Technology (VCQ), Boltzmanngasse 5, A-1090 Vienna, Austria}


\date{\today}% It is always \today, today,

\maketitle

\onecolumngrid

As mentioned in the main text, the same computational analysis as above was performed for an additional range of TMDs. Specifically, we examined MoS$_2$, MoSe$_2$, MoTe$_2$ and WSe$_2$. Their indices of refraction and densities are displayed in Table \ref{allTmdtable} below.
\begin{table}[h!]
\centering
\caption{\label{allTmdtable}Index of refraction $n$, extinction coefficient $k$, and density of different TMD's at $\lambda = 1550$ \si{nm}.}

\begin{tabular}{l c @{\hspace{1cm}} c c} % added 1cm space between n and k
\toprule
Material & $n$  & $k$  & Density (g/cm$^3$) \\
\midrule
WSe$_2$  & 3.99&$<$ \num{6.0d-9} \cite{TMDref_indexes}                                   &      9.32      \cite{handbook_of_chem}   \\
MoTe$_2$ &  4.84   &     $<$ \num{3.0d-8} \cite{TMDref_indexes}                                   &      7.99      \cite{MoTe2_density}             \\
WS$_2$   & 3.79 & $<$ \num{2.83d-10} \cite{WS2n}      & 7.50 \cite{densityws2} \\
MoSe$_2$ &  4.22 &  $<$ \num{1.0d-7} \cite{TMDref_indexes}                                      &     6.99   \cite{handbook_of_chem}                  \\
MoS$_2$  & 3.76 & $<$ \num{2.37d-11} \cite{TMDref_indexes} & 5.06 \cite{handbook_of_chem} \\
\bottomrule

\end{tabular}
\end{table}

\subsection*{Quadrupolar force}
\begin{figure}[h]
\includegraphics[width=0.95\linewidth]{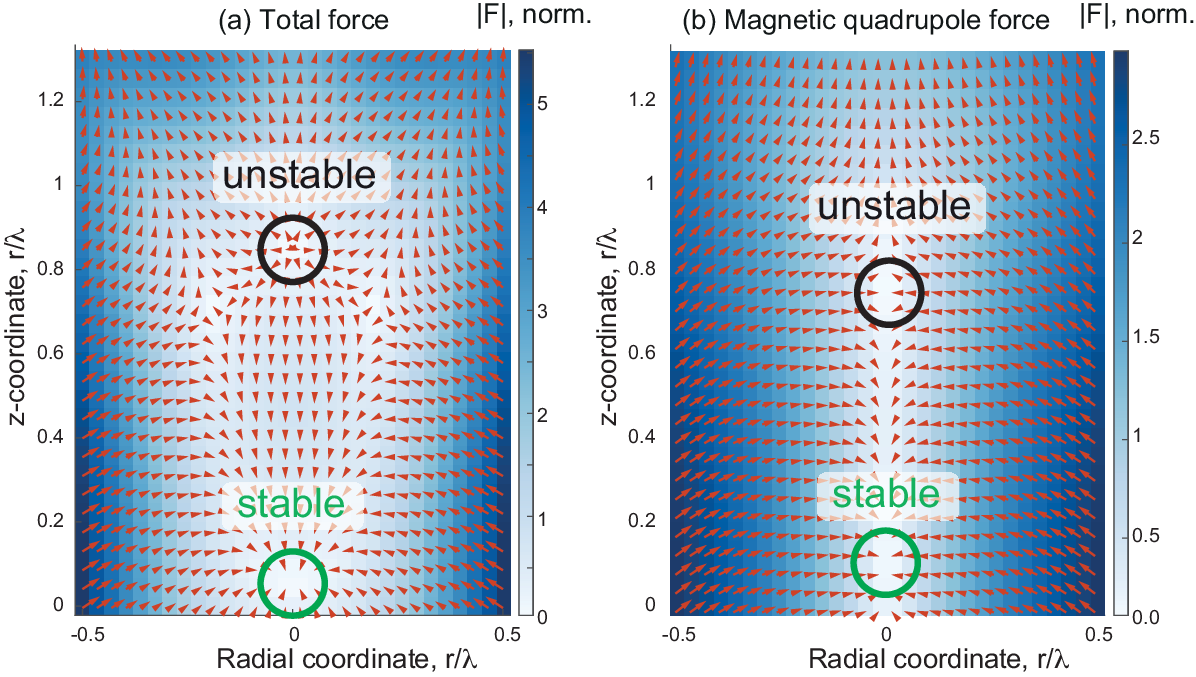}
\caption{\label{fig_MQ_force}(a) Distribution of optical forces in the beam cross-section with account of all multipole components and (b) with MQ contribution only.}

\vspace{-7pt}
\end{figure}

Figure~\ref{fig_MQ_force} compares the spatial distribution of the optical force in the transverse beam cross-section for the full multipolar response (a) and for the magnetic quadrupole (MQ) contribution alone (b). In both cases, the force field exhibits a stable equilibrium point along the beam axis at lower $z$, where the force vectors converge toward the intensity minimum, and an unstable region at higher $z$ characterized by outward-directed forces. The MQ-only contribution reproduces the qualitative structure of the trapping landscape, including the existence and approximate location of the stable point, confirming its dominant role near the resonance. However, quantitative differences in force magnitude and stability boundaries are evident, indicating that the full trapping behaviour arises from the interference of multiple multipolar components rather than the MQ term alone.
%===========Figure=====================
%===========Figure=====================

\subsection*{Equilibrium position and trapping frequency}
\begin{figure}[h]
\includegraphics[width=1\linewidth]{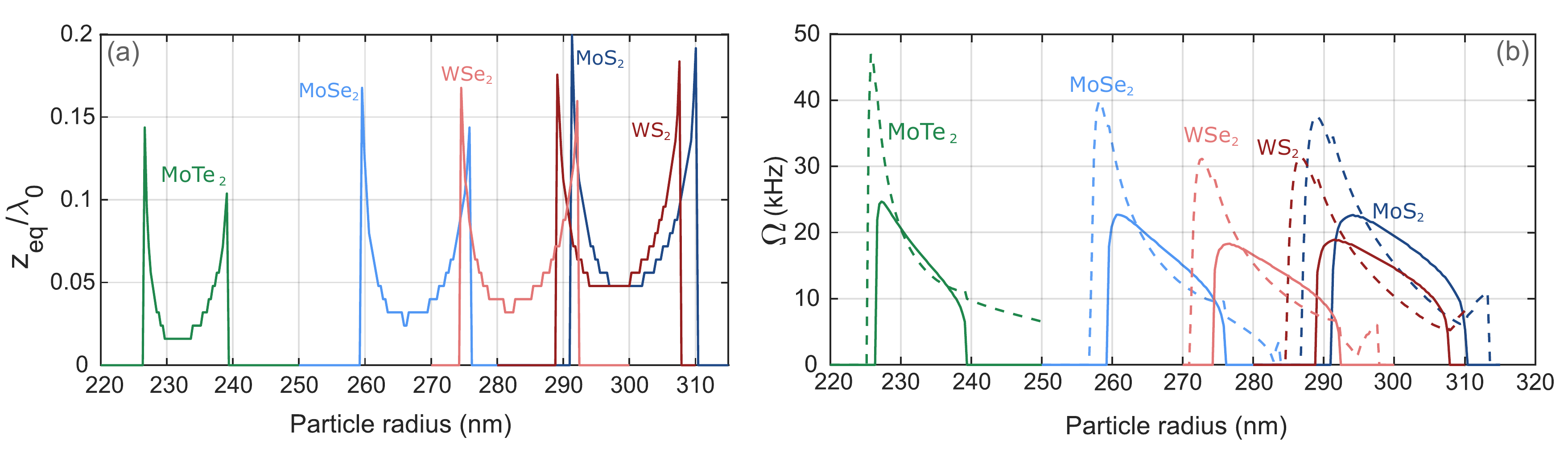}
\caption{\label{fig1sup}(a) Equilibrium position z$_{\text{eq}}$ of trapped TMD particles, normalized by the trapping wavelength $\lambda_0 = 1550\,\mathrm{nm}$, as a function of particle radius. (b) Corresponding axial trapping frequency (solid lines) and radial trapping frequency (dashed lines) for different TMD materials. For particle radii at which no stable equilibrium position exists, the z trapping frequency is set to zero.}

\vspace{-7pt}
\end{figure}

Due to the differing refractive indices of the various TMDs, the particle radius range over which stable trapping occurs varies from one material to another. Similarly, the corresponding trapping frequency also varies for different TMD's.  Fig \ref{fig1sup} shows the equilibrium positions and the trapping frequency for the different TMD's, for different particle radii. A similar trend can be observed across all TMDs.
%===========Figure=====================
%===========Figure=====================

\subsection*{Recoil rate}
\begin{figure}[h]
\includegraphics[width=0.65\linewidth]{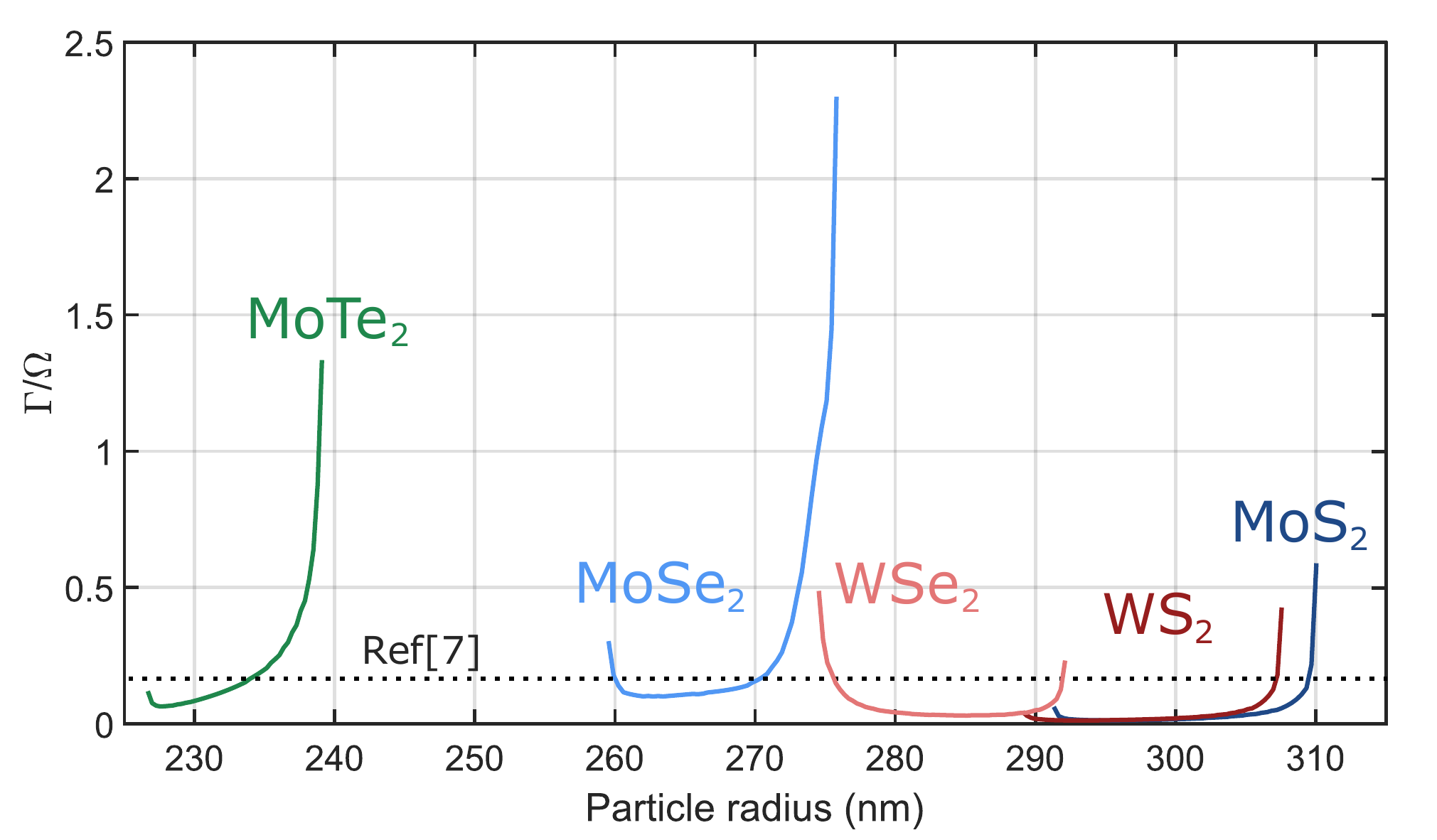}
\caption{\label{fig2sup}The dependence of the recoil rate $\Gamma$ to the trap frequency $\Omega_z$ ratio on the particle radius for the different TMDs. Added is a black-dotted line which corresponds to a SiO$_2$ particle of radius R =
73 nm in a standard bright trap \cite{lorenzocontrol}.}

\vspace{-7pt}
\end{figure}

Figure \ref{fig2sup} shows the dependence of the ratio between the recoil heating rate $\Gamma$ and the trap frequency $\Omega_z$ on the particle radius for different TMD materials. The curves highlight the strong material- and size-dependent behaviour of $\Gamma/\Omega_z$. The black dotted line indicates the corresponding value for a SiO$_2$ particle with radius $R = 73$ nm trapped in a conventional bright optical trap of lower power \cite{lorenzocontrol}, serving as a reference for comparison. One can note that an enhanced ratio $\Gamma/\Omega_z$ is not obtained uniformly across particle sizes/materials. Instead, the improvement occurs only within specific ranges of particle radii, where resonant optical effects significantly modify the light scattering. 
%===========Figure=====================
%===========Figure=====================
\newpage

\subsection*{Internal temperature}
\begin{figure}[h]
\centering
\includegraphics[width=1\linewidth]{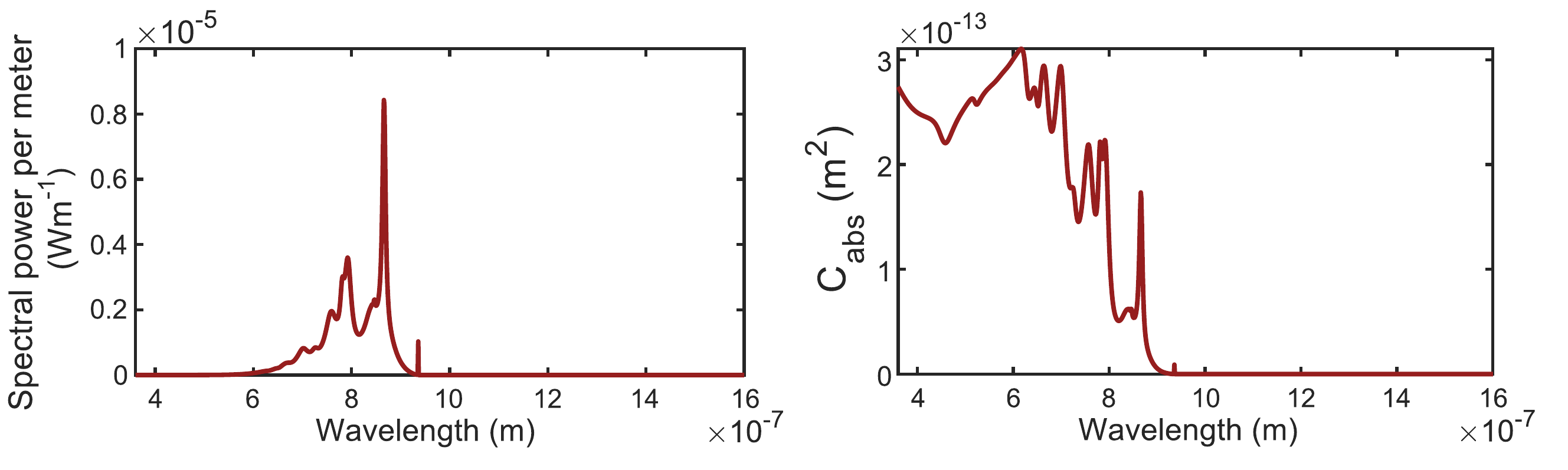}
\caption{\label{emission_spectra_sup}Left side: Spectral power per unit meter of a trapped particle in the dark trap at its equilibrium temperature. Right side: absorption (emission) cross-section of a $300\si{ nm}$ WS$_2$ particle.}

\vspace{-7pt}
\end{figure}

Here, we provide the spectral characteristics underlying the heating and radiative cooling mechanisms discussed in the main text. We evaluate the absorption cross-section and the emitted power per meter of a WS$_2$ particle trapped in the dark trap. Figure \ref{emission_spectra_sup} (right side) shows the absorption cross-section $C_{\mathrm{abs}}$ of the particle as a function of wavelength. Pronounced peaks are observed, corresponding to the Mie resonances supported by the particle. One can also observe how, as the imaginary part of the refractive index goes to zero, so does the absorption.

To quantify thermal emission, we consider the spectral radiated power per meter $P_m$ in \ref{emission_spectra_sup} (left side). This is obtained by \cite{metaatoms}: 
\begin{equation}
P_m = C_{\mathrm{abs}}(\lambda)\times B_\lambda(T),
\end{equation}
where $B_\lambda(T)$ is the Planck spectral radiance at temperature $T$. The total radiated power is then obtained by integrating $P_m$ over the wavelength range.

%===========Figure=====================
%===========Figure=====================

\begin{figure}[h]
\centering
\includegraphics[width=1\linewidth]{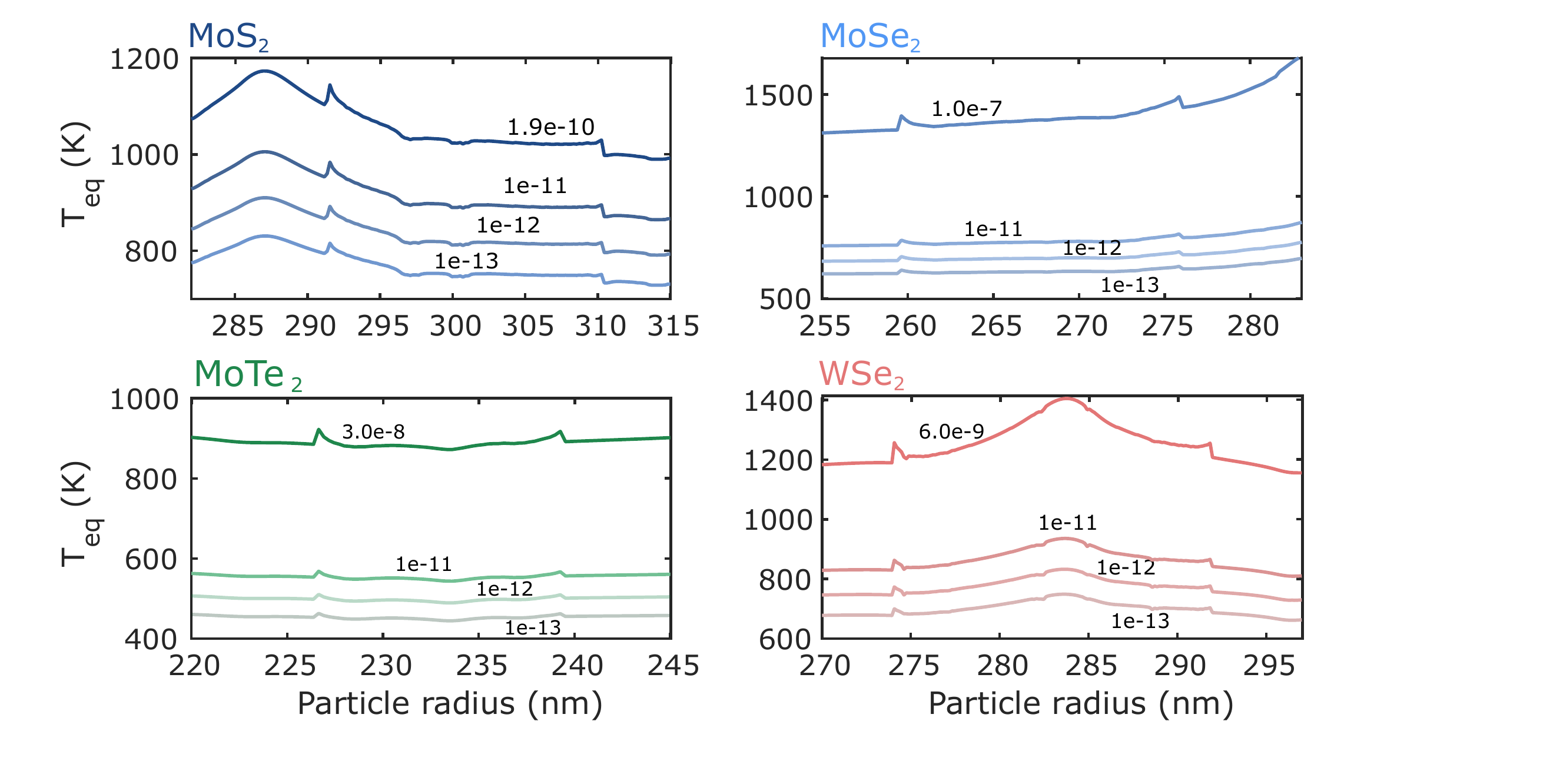}
\caption{ \label{alltemps}Equilibrium temperature T$_{\text{eq}}$ of different material trapped particles for different extinction coefficients $k$ plotted against particle radius. The coloured text on the top left of each plot indicates the specific TMD. The text at the top of each curve indicates the corresponding extinction coefficient. The melting (decomposition) temperatures of the materials are the following: MoS$_2$: 2650 K \cite{melting_mos2}, MoSe$_2$: $>$ 1470 K \cite{handbook_of_chem} , MoTe$_2$ : 1000 K \cite{mote2_temp} , WSe$_2$: 1470 K \cite{melting_wse2}}.
\vspace{-7pt}

\end{figure}

Figure \ref{alltemps} shows the equilibrium temperature T$_{\text{eq}}$ of different optically trapped TMD particles as a function of particle radius for different extinction coefficients. While the absolute temperature values and their radius dependence vary among TMDs due to differences in absorption efficiency, the overall trend of increasing equilibrium temperature with extinction coefficient is common to all materials.
%===========Figure=====================

\putbib
\end{bibunit}
\end{document}